\documentclass{PoS}

\title{Galactic Cosmic Rays - Theory and Interpretation}

\ShortTitle{GCR Theory}

\author{\speaker{Luke O'C. Drury}\\
        Dublin Institute for Advanced Studies\\
        School of Cosmic Physics\\
        31 Fitzwilliam Place\\
        Dublin 2\\
        Ireland
        E-mail: \email{ld@cp.dias.ie}}

\abstract{The arguments surrounding the Galactic component of the cosmic rays, the energy budget, questions of composition, spectral features, anisotropy, sources etc, will be critically examined. We are moving into a new phase in the study of the Galactic cosmic rays where it is becoming clear that precision measurements are revealing new, and in some cases unexpected, features which are forcing us to develop more sophisticated models for their production and propagation. The fundamental concepts however appear to be quite solid and have changed remarkably little in the more than fifty years since Ginzburg and Syrovatskii's classic "The Origin of Cosmic Rays".
}

\FullConference{35th International Cosmic Ray Conference --- ICRC2017\\
		10--20 July, 2017\\
		Bexco, Busan, Korea}

\usepackage{aas_macros}

\begin{document}

\section{Introduction}
Framing the theory and interpretation of cosmic rays as a corner-stone of high-energy astrophysics can largely be traced back to the influence of the 1964 monograph "Origin of Cosmic Rays" by Ginzburg and Syrovatskii \cite{1964ocr..book.....G}. This genuinely seminal book transformed cosmic ray physics from being the poor relation of particle physics into the important branch of high-energy astrophysics which it is today.  Looking back it is remarkable how prescient Ginzburg and Syrovatskii were.  They made a number of key points:
\begin{enumerate}
\item {Cosmic Rays are essentially a Galactic phenomenon, neither Solar nor cosmological, and their study is properly part of astrophysics.}

\item{The cosmic ray population that we observe locally in the Solar neighbourhood is a representative sample of that throughout most of the Galaxy; what one might call a cosmic ray Copernican principle.}

\item{Cosmic Ray propagation is by spatial diffusion in an extended magnetised halo surrounding and anchored in the gas disc of the Galaxy.}

\item{The cosmic ray energy budget points firmly to supernovae as the power source  (it "jumps off the page" as they vividly put it).}

\item{The great potential of gamma-ray and neutrino astronomy is emphasised - a remarkable feature for a book published over fifty years ago!}

\end{enumerate}

Most of this is still largely true.  While the Galactic nature of the bulk of the cosmic rays is nowadays taken for granted it should be remembered that Alfv\'en, Richtmeyer and  Teller \cite{1949PhRv...75..892A, 1950PhRv...77..375A} were arguing for a purely Solar origin as late as the 1950s, and that on the other hand Hoyle and Burbidge \cite{1964PPS....84..141B} were arguing that the entire universe was filled with a uniform sea of cosmic rays, leading to a long running controversy between Ginzburg and Burbidge (see e.g. \cite{1971CoASP...3..140B} where Burbidge writes "in modern times the extragalactic theory has been continuously attacked by Ginzburg and Syrovatskii, and their arguments against this hypothesis have been repeated quite uncritically by others") as described in Longair's book \cite{1981heaa.book.....L}. Certainly the energetically dominant component of the cosmic rays, the mildly-relativistic nuclei with energies around a GeV per nucleon, are a Galactic phenomenon whereas the ultra-high energy cosmic rays are probably extra-galactic although the transition, probably in the EeV region, is uncertain (see review by Andrew Taylor at this conference). As we will see the Galactic cosmic rays (hereafter GCR) do fill the galactic disc rather uniformly and isotropically, in fact surprisingly so.  The transport of the GCR does have a strong diffusion component, but is almost certainly more complicated than in the Ginzburg pure diffusion model. Perhaps most significantly, the energy argument has not changed and is still compelling evidence for Supernovae as the ultimate power source driving GCR acceleration. And finally of course, their predictions of a bright future for Gamma-ray astronomy have been realised and we are on the verge of opening up a genuine neutrino astronomy. 

\section{Distribution within the Galaxy}

Ginzburg and Syrovatskii had to rely mainly on radio synchrotron observations of our own and other spiral galaxies to infer that the entire disc was filled with relativistic cosmic ray electrons which also extended some distance above and below the disc to form a "halo", and then assume that the same held for the nuclear species.  They pointed however to the great potential of gamma-ray astronomy (and in principle neutrino astronomy) to address this question.  The dominant source of the diffuse gamma-ray emission in the Galaxy above about a GeV is the decay of neutral pions produced by hadronic interactions between cosmic ray nuclei and the atoms of the interstellar gas.  Thus the gamma-ray emissivity traces the product of the cosmic ray intensity and the local gas density (of course if one wants to do a proper quantitative analysis it is important also to include inverse Compton contributions from the cosmic ray electrons as well as bremsstrahlung).  Even a cursory comparison of the Femi-LAT all sky maps above a GeV with the distribution of dust and gas in the Galaxy as revealed by e.g. the Planck data immediately shows the two look almost identical from which it follows that the cosmic ray flux throughout the Galaxy must be pretty uniform.  In fact detailed studies show that the bulk of the disc must be filled with cosmic rays very similar (in intensity and energy spectrum) to those observed locally.  Only in the immediate neighbourhood of the Galactic centre is there evidence for a harder spectrum and higher intensity, and while there is a falling off towards the outer Galaxy the radial gradient is surprisingly small \cite{2015ARA&A..53..199G}.  The radial gradient, together with the distinctly lower emissivity observed in the Magellanic clouds (a classic test proposed by Ginzburg \cite{1972NPhS..239....8G}), clearly show that the bulk of the lower-energy cosmic rays are a Galactic phenomenon coming from sources within our Galaxy.

What the Fermi-LAT observations have also revealed, and what nobody anticipated, is the presence of two large extended structures above and below the Galactic plane, the so-called Fermi bubbles.  These sharp-edged and limb-brightened features with a hard energy spectrum are not well understood, but clearly point to some form of past nuclear activity in our own Galaxy, see e.g. \cite{2014ApJ...793...64A, 2015ApJ...808..107C} and references therein.  

\section{Anisotropy}

The arrival direction distribution of the charged cosmic rays is remarkably isotropic, with typical anisotropies of less than $10^{-3}$ over the energy range thought to be Galactic (see e.g. \cite{2014arXiv1407.2144D} and references therein). The reason is fairly obvious; magnetic fields deflect particles and are very effective in isotropising charged particle distributions, so the near isotropy of the GCR points to strong scattering by Galactic magnetic fields, however the quantitative details are complex.  Over the last decade the observational situation has greatly improved thanks to a series of experiments starting with Milagro \cite{2008PhRvL.101v1101A} and followed by HAWC, IceCube, Argo-YBJ, EAS-Top, Tibet-ASg etc which have greatly improved our knowledge of the arrival direction distributions of GCRs at PeV energies (a classic example of one man's background being another man's signal - for the most part these are gamma-ray and neutrino observatories!), see e.g. \cite{2017arXiv170803005T}. These have revealed a wealth of detail, including both small-scale and large-scale features.  The existence of the small-scale (10 degrees or so on the sky) structure was initially a surprise because in simple diffusion models only a dipole anisotropy is expected, but is now I believe reasonably well understood.  The key point is that the angular distributions observed in any particular instantiation of a random magnetic field are not the same as those in the ensemble average \cite{2012PhRvL.109g1101G, 2017PrPNP..94..184A} and can in principle be used as probes of the local ISM field structure \cite{2017ApJ...835..258G, 2017arXiv170201001G} as well as Heliospheric magnetic (and electric  \cite{2013arXiv1305.6752O}) fields \cite{2016JPhCS.767a2027Z}. The large scale structure possibly hints at interesting local sources such as the Vela SNR, see e.g. \cite{2016PhRvL.117o1103A}.  The bottom line remains however that the low level of anisotropy, and in particular its rather weak energy dependence, is a strong constraint on propagation and is in clear tension with simple models based on pure diffusion (and the unphysical leaky box, which is best thought of as a one-zone approximation to the diffusion model).

\section{The Galactic cosmic ray luminosity}

The question of how much power is required to maintain the observed cosmic ray population in the Galaxy, or to put it another way what is the cosmic ray luminosity of the Galaxy $L_{\rm GCR}$, is, as recognised by Ginzburg and Syrovatskii, a key issue.  The conventional estimate is $10^{41}\rm erg\,s^{-1}$ or in SI units $10^{34}\,\rm W$. Ginzburg and Syrovatskii estimated back in 1964 a value one third of this, $0.3\times 10^{34}\,\rm W$ whereas Drury, Markiewicz and V\"olk in 1989 \cite{1989A&A...225..179D} suggested a high value of $3\times 10^{34}\,\rm W$.  More recently Strong et al using the Galprop code quoted the remarkably precise value of $(0.7\pm0.1)\times 10^{34}\,\rm W$, but it is clear that the error here is the statistical error in the measurements and that there is a much larger systematic error associated with the choice of propagation model.

The basic power estimate is quite robust and it is worth explaining in some detail how it arises.  The key point is that the local energy density of the GCRs, and their so-called "grammage" are both very well constrained by the observations at a few GeV per nucleon where we have excellent data.  The "grammage" is simply the amount of matter traversed by the GCRs before reaching the solar neighbourhood and is fixed by observations of the ratios of secondary spallation nuclei to primary source nuclei, for example the Boron to Carbon ratio.  This unambiguously shows that the GCR we observe at Earth have traversed a grammage of $g\approx 5 \rm\, g\, cm^{-2}$ at a few GeV per nucleon.

Let us now assume that the GCR have been confined for a time $\tau$ in a volume of size $V$ containing a target mass $M$, then
\begin{equation}
g\approx {\tau c M\over V},
\end{equation}
and if their energy density is $\mathcal E_{\rm GCR}$, then the Galactic GCR luminosity is just
\begin{equation}
L_{\rm GCR} \approx {\mathcal E_{\rm GCR} V\over \tau}
\end{equation}
from which it trivially follows that 
\begin{equation}
L_{\rm GCR} \approx \mathcal E_{\rm GCR} {c M\over g}.
\end{equation}
Taking fairly standard values of a local GCR energy density $\mathcal E_{\rm GCR} \approx 1.0\,\rm eV\, cm^{-3}$, a total interstellar gas mass in the Galaxy of $5\times 10^{9} \rm M_\odot$ and $g\approx 5 \rm\, g\, cm^{-2}$ gives 
$L_{\rm GCR} \approx 10^{34}\rm\, W$.  

It is important to note that this does not depend on the exact value of the confinement time $\tau$ or volume $V$, all that matters is their ratio.  In particular the argument is independent of cosmic ray ages inferred from radioactive secondaries such as $^{10}\rm Be$ etc.  It does of course rely on the cosmic ray Copernican assumption, that there is nothing special about the solar neighbourhood, but this is well supported by the gamma-ray observations as discussed above. It also assumes that the cosmic ray energy does not change significantly during propagation, that they more or less uniformly sample all the interstellar gas (in particular that they penetrate molecular clouds), and it only determines the luminosity in mildly relativistic particles around a GeV per nucleon.  All of these reflect issues to do with the choice of cosmic ray propagation model.

Historically the oldest propagation model, and one that refuses to die even though it is clearly unphysical, is the so-called leaky box model with a fixed $V$ and an energy dependent escape time $\tau$.  This is probably best thought of as a useful approximation to a more physical Ginzburg-type spatial diffusion model.  At the phenomenological level one could however equally well consider both the volume $V$ and escape time $\tau$ to be energy dependent, and indeed this "expanding leaky box" can be shown to be an approximation to more sophisticated dynamical outflow models, see \cite{1996A&A...311..113Z} and more recently \cite{2017MNRAS.470..865R}.  Putting some of the energy dependence into the effective volume has the  advantage of making the anisotropy data easier to understand - the larger confinement volume at high energies lowers the expected anisotropy.

Observationally it is clear from the secondary to primary ratios that $\tau/V$ is a decreasing function of energy, and that thus the cosmic ray production spectrum must be harder than that observed.  Theory tends to favour quite hard production spectra close to $N(E)\propto E^{-2}$ whereas observations seem to favour rather softer injection spectra closer to $E^{-2.3}$.  This has obvious implications for the energy budget and the high estimate of \cite{1989A&A...225..179D} is a consequence of assuming a hard production spectrum (and subsequently more rapid escape or greater dilution in a larger volume to soften the spectrum) so that more energy has to be put into producing the higher energy particles.

In addition to this uncertainty at high energies, at low energies the assumption that particles propagate without significant energy change is questionable.  There must be some level of second order Fermi acceleration associated with the scattering of the particles by the magnetic field unless the field is strictly magnetostatic.  In reality of course the field irregularities responsible for the scattering are expected to be moving with characteristic velocities of order the Alfv\'en speed and thus the spatial diffusion should be accompanied by a certain amount of momentum-space diffusion.  For historical reasons this effect is usually called "re-acceleration" and is incorporated in propagation codes such as Galprop where it is typically adjusted to give a better fit to the low-energy Boron to Carbon data, e.g. \cite{2016ApJ...824...16J}.  Recently \cite{2017A&A...597A.117D} Andy Strong and I have evaluated how much energy is actually transferred to the cosmic rays through this process if the usual parameter values are used and shown that it implies an uncomfortably high level of re-acceleration power, perhaps as high as $5\times 10^{33}\,\rm W$ or half the cosmic ray luminosity of the galaxy!  While not absolutely impossible this seems implausible and we note that advection in an outflow or a directed wave field (see e.g. \cite{2015A&A...583A..95A}) can give an equally good fit to the low-energy data.

To summarise the energetics, we can safely assume that the total Galactic cosmic ray luminosity is
\begin{equation}
0.3\times 10^{34}\,\rm W < L_{\rm GCR} < 3\times 10^{34}\,\rm W
\end{equation}
with the uncertainty reflecting systematic uncertainties in the propagation models rather than statistical uncertainties in the data.
As much as half of this may possibly come from re-acceleration if the standard Galprop parameter fittings and model are used.

Now the key point made long ago by Ginzburg and Syrovatskii is that the mechanical power injected by Galactic supernovae is of order $P_{\rm SNe} \approx 10^{35}\,\rm W$ and, apart perhaps from the Galactic centre, there is no other plausible source of sufficient energy to drive the acceleration.  Pulsars and OB winds may make a minor contribution at the 10\% level, but are not as powerful.  The fact that the power available from supernova explosions is precisely that required to produce the GCR luminosity if the accelerator works at about 10\% efficiency, and that nothing else seems to be available is a powerful argument for saying that the energy powering the GCR must ultimately come from supernovae.  This argument remains as valid today as it was in 1964 (and indeed the idea goes back even further to the discovery of supernovae by Baade and Zwicky in 1934 \cite{1934PhRv...46...76B}).  

Assuming that the most plausible energy source to drive the acceleration is the explosion energy of supernovae, the rather high efficiency required (of order 10\%) implies that it cannot take place directly in the explosion itself because there would then be a major problem with adiabatic losses in the subsequent expansion of the remnant.  It follows that it must be mediated through the shocks associated with the remnant, or possibly to some extent the general ISM turbulence (see fig 1).  
\begin{figure}
\begin{center}
\includegraphics[width=\textwidth]{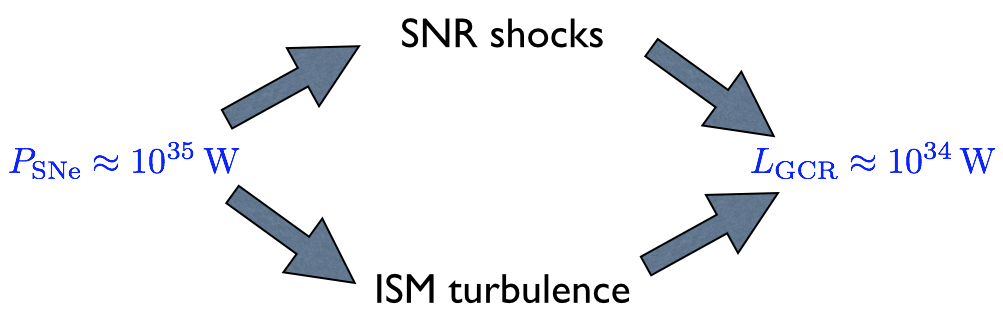}
\label{Fig1}
\caption{The putative energy flow from SNe to GCRs}
\end{center}
\end{figure}

Turning to the Galactic centre we know that it contains a supermassive black hole with a mass of $4\times10^{6}\,\rm M_\odot$ \cite{2009ApJ...692.1075G}. The natural power scale associated with accretion onto a black hole is the Eddington luminosity, which in the case of the Galactic centre is $5\times 10^{37}\,\rm W$.  Clearly, and perhaps fortunately, it is massively sub-luminous at the present time, but there is evidence for significant time variability, e.g. \cite{2016MNRAS.463.2893W}, and thus it could easily make a significant time-averaged contribution if it flickers on and off as many AGN do.  The recent evidence from the HESS collaboration of particle acceleration to PeV energies in the Galactic centre is very exciting in this regard \cite{2016Natur.531..476H}. Combined with the recent Fermi-LAT observations showing a harder spectrum in the Galactic centre region, e.g. \cite{2017PhRvL.119c1101G}, and the evidence of the Fermi bubbles, it is hard to escape the conclusion that there is a lot of high-energy particle acceleration associated with the centre of our Galaxy.  Whether this is related to episodic accretion onto the black hole, or star-burst activity in the central molecular zone remains open.  It should be noted that exciting as the HESS observations are, they imply a luminosity in PeV particles from the Galactic centre that is currently quite modest at around $10^{30} -10^{31}\,\rm W$ and certainly not enough at the moment to make a major contribution to the general GCR population.  Of course the Fermi bubbles suggest much more violent activity in the past, and it is not impossible that in the "knee" region we see a significant contribution from past Galactic centre activity.

\section{Charge resolved energy spectra}

Even if little has changed in regard to the overall energy and power considerations, the last few years have seen exciting new developments in our knowledge of the particle energy spectra.  Until recently the paradigm was that all the primary GCR nuclei, after correcting for solar modulation and spallation during propagation, were essentially just one feature-less power law between a few GeV per nucleon and the "knee" at around $3\times 10^{15}\,\rm eV$.  It is now clear, thanks to precision measurements by first Pamela \cite{2011Sci...332...69A} and now AMS02 \cite{2017NCimC..39..313I, 2015PhRvL.114q1103A, 2015PhRvL.115u1101A} building on earlier work by the ATIC \cite{2006AdSpR..37.1950A, 2006astro.ph.12377P}and CREAM \cite{2010ApJ...714L..89A, 2011ApJ...728..122Y} collaborations, that this is not the case.  These experiments have resulted in two important discoveries:
\begin{enumerate}
\item {The proton spectrum is distinctly softer than that of Helium (and possibly other heavy elements) at all energies.}
\item{Both spectra show a break and a spectral hardening at around a rigidity of $200\,\rm GV$.}
\end{enumerate}
Also for the first time we have low-energy measurements from the Voyager spacecraft showing the low-energy interstellar spectrum of GCR protons largely free of Solar modulation \cite{2013Sci...341..150S, 2015ApJ...815..119V}. In addition to these exciting developments concerning the nuclear species, the precise measurements of the positron, electron and anti-proton components are also throwing up new ideas and challenges, see e.g. \cite{2017PhRvD..95f3009L} for an unconventional take on the anti-particle data. 

The harder helium spectrum has the interesting consequence that by the time one gets to the "knee" energies it dominates hydrogen in the all-particle energy spectrum (though not in energy per nucleon or rigidity).  Thus the "knee" in the all-particle spectrum at $3\times10^{15}\,\rm eV$ is actually predominantly a Helium and CNO knee, and it is possible that the proton spectrum cuts off significantly before this as has been suggested by the Tibet ARGO-YBJ experiment \cite{2014NIMPA.742..241M}.  (This is reminiscent of the very old Grigorov claims of a deficit of protons at the "knee" although Grigorov's results \cite{1972spre.conf.1617G} were almost certainly due to splash-back from his calorimeter leading him to misclassify protons as heavy nuclei - it is interesting however that he may well have been partially right for the wrong reason.)

It is hard to see how the discrepancy in slope between the proton and helium spectra can be a propagation effect and it is thus almost certainly a source effect.  Either there are distinct populations of sources, hydrogen rich ones producing rather softer spectra and helium rich ones producing systematically harder ones, or it is an effect of source evolution, with the hydrogen and helium being preferentially injected and accelerated at different times as the source (presumably a SNR) evolves (see e.g. \cite{2017arXiv170702744H}).  In some ways this is not too surprising.  We know that supernovae come in a wide variety of types and often explode in complex and highly structured environments.  It would be astonishing were they all to produce identical populations of accelerated particles throughout their evolution, and it seems quite natural that Helium, with its high ionisation potential and enhancement in the winds of high-mass stars at the end of their evolution, should be preferentially associated with very strong shocks and the early phases of evolution of core-collapse supernovae where one might expect harder spectra.  It will be important in the next few years to see whether other abundant elements also show similar variability in spectral slope and whether a consistent pattern emerges.

The break at rigidity of $200\,\rm GV$ or so on the other hand could reflect either a source effect (concave spectra are a generic prediction of both linear and nonlinear acceleration theories) or a propagation effect.  In principle the question should be answered by precise measurements of secondary spallation nuclei.  If it is a pure propagation effect, the break in the secondary spectra should be twice that in the primary spectra, and this should be evident in the spectrum of a secondary species such as Li and also as a break in secondary to primary ratios such as B/C. Unfortunately the data presented to date do not seem to allow a clear answer to this question (see \cite{2017arXiv170609812G} for a different perspective), but it should be clarified in the near future with the promised release of improved AMS02 results as well as results from new experiments such as DAMPE and ISS-CREAM.  Such effects are actually quite natural in more sophisticated propagation theories a good example being \cite{2015A&A...583A..95A} where they make the point that their model also naturally fits the low-energy Voyager data.

\section{Chemical composition}

The relative abundances of the different nuclear species in the GCR provide very important information about their origin and propagation.  As noted by Ginzburg and Syrovatskii "It can be said with certainty that any theory of the origin of cosmic rays cannot expect serious success unless it rests on a detailed analysis of the observed composition of primary cosmic radiation".  By relative abundance we normally mean the ratio of the flux in one species to some reference species measured at a fixed energy per nucleon.  It would probably make more sense to do this at fixed rigidity, but it makes no real difference except in the case of hydrogen.  Implicitly this relies on the fact that all nuclear species have very similar spectra so that there is only a weak dependence on the measurement energy, but the recent discovery of the differences between the hydrogen and helium spectra shows that some care is needed here and the composition is clearly energy dependent.  {\it A fortiori} statements about the relative numbers of electrons and protons in the GCR should be treated with great caution, they have very different spectra and it is not clear how one should make such a comparison. Leaving aside these minor caveats, we now have excellent data for all nuclei up to Fe (most recently from the ACE collaboration \cite{2005NuPhA.758..201I}), good data for some of the trans-Iron elements (see the recent results from the TIGER collaboration, e.g. \cite{2016ApJ...831..148M}), and evidence for the presence of all nuclei including the actinides in the GCR \cite{1998NIMPB.145..409W, 2012ApJ...747...40D}. 

The data show clearly that the composition of the GCR, after correction for spallation during propagation, is relatively normal and quite close to that of the solar system (which is commonly used as a standard reference composition).  All the main nucleosynthetic groups are present in proportions similar to those in the solar system.  This is interesting because it implies that we are not looking at the ejecta of any one class of supernovae, instead we need a mixture similar to that in the general interstellar medium (which is thought to be very similar to the solar system composition). There are however some significant differences in the GCR composition as compared to the solar system.  The most striking is that heavy elements are definitely over-abundant, and that there appear to be two parameters at work.  On theoretical grounds one expects that there should be a mass to charge ratio dependent fractionation, with higher initial rigidity ions being preferentially injected, but this alone is not enough to explain the data \cite{1981ICRC....2..269C} and there needs to be a second effect which phenomenologically is related to the chemistry or the outer electronic structure of the neutral atom (first ionisation potential, volatility etc).  This is clearly telling us something about both the source material that is being accelerated and the initial injection and acceleration process.  Remarkably the heavy and refractory elements that are overabundant in the GCR, throughout most of the interstellar medium are strongly depleted in the gas phase and condense out as dust grains.   It is therefore natural to ask whether it is possible to imagine mechanisms whereby dust could contribute to the GCR source composition and indeed just such a quantitative model was sketched out in \cite{1997ApJ...487..182M, 1997ApJ...487..197E} building on earlier suggestions \cite{1980MNRAS.193..723E, 1981IAUS...94..361C}.   The bottom line is that rather standard shock acceleration in a dusty medium of normal composition appears to be well able to fit the GCR compositional data.  Perhaps the most convincing feature of the model is the case of oxygen which is well fitted once one allows for the fact that about 10\% of the interstellar oxygen is trapped in the grains (which are mainly silicates and metal oxides; note that this is an argument for chemistry and not just atomic physics as the second parameter).

Of course using the solar system composition as a reference is more a matter of convenience than anything else, and the TIGER collaboration in a series of recent papers have been using instead a composition enriched in mass-loss from massive stars to model what might be the composition in a superbubble.  This appears to organise the data better and makes sense in that most core-collapse supernovae probably occur in such an environment \cite{2009ApJ...697.2083R, 2016ApJ...831..148M}.  It would be well worth doing a more detailed study of the gas/dust injection problem using modern computational resources to verify that the model is indeed qualitatively correct.  A first step in this direction is the {\it ab initio} calculation of gas phase fractionation in \cite{2017arXiv170408252C}.

In addition to the chemical composition we also have some isotopic information.  The best established effect is the clear excess of $^{22}$Ne generally interpreted as pointing to a contribution from Wolf-Rayet star winds \cite{1982ApJ...258..860C, 2012A&A...538A..80P}.  The remarkable recent detection of live $^{60}$Fe \cite{2017APS..APR.M4009B} in the GCR points to the presence of some freshly synthesised material (less than a few million years old) while on the other hand the absence of $^{59}$Ni (which decays by k-shell capture and is thus stable when fully stripped) has been interpreted as indicating an interval between nucleosynthesis and acceleration of at least $10^5$ years \cite{1999ApJ...523L..61W} (but see \cite{2016A&A...588A..86N} for a dissenting view). This is consistent with the actinide composition which also hints at a generally old composition contaminated by a small amount of freshly synthesised  material \cite{2012ApJ...747...40D}.

\section{Acceleration sites and mechanisms}

Thus the energy to power the bulk of the GCR acceleration is almost certainly from supernovae, and the matter that is accelerated appears to be a well-mixed rather normal dusty interstellar medium with some fresh nucleosynthetic contamination.  But how and where does the acceleration actually happen? The most plausible suggestion goes back to the ideas of \cite{1977DoSSR.234.1306K, 1978MNRAS.182..147B, 1978ApJ...221L..29B, 1977ICRC...11..132A}, a first-order Fermi acceleration process now generally referred to as diffusive shock acceleration (after \cite{1983RPPh...46..973D}) which locates the site of the acceleration in the strong shocks associated with supernova remnants.   Stochastic acceleration by turbulence (the classical second-order Fermi mechanism) remains a viable possibility, but is generally very much slower, requires a separate injection process, and there are many competing channels for the dissipation of turbulence.  Diffusive shock acceleration has the great advantage of being able to accelerate charged particles directly out of the tail of the shock-heated ion distribution with high efficiency as confirmed recently by detailed numerical simulations \cite{2014ApJ...783...91C, 2014ApJ...794...46C, 2014ApJ...794...47C} and naturally produces hard power-law spectra approximating to $E^{-2}$.  Worries about the maximum attainable particle rigidity \cite{1983A&A...125..249L} have been somewhat alleviated by the introduction of magnetic field amplification \cite{2000MNRAS.314...65L, 2004MNRAS.353..550B, 2014MNRAS.444..365D}.  Any strong shock in the ISM should accelerate particles in the same way and thus there may be a contribution from the shocks associated with OB winds, but these do no have enough power to explain the bulk of the GCR. Similarly pulsars may, and in my opinion probably do, contribute to the electron and especially the positron components of the GCR, but it is difficult to see how they on their own could reproduce either the power or the composition of the GCR.

In addition to the Fermi processes which tap into differential motion to drive particle acceleration, there is one other possible source of energy to drive acceleration and that is the energy density of the magnetic field released through magnetic reconnection.  While there is no doubt that magnetic reconnection does drive particle acceleration in a number of astrophysical environments (Solar flares being the paradigmatic example) it is difficult to see how it could produce the Galactic luminosity or explain the composition of the GCR.  The other problem with magnetic reconnection is that the theory is much harder and less specific than in the case of diffusive shock acceleration.

Thus the "best bet" remains as it has been for the last few decades: the GCR are produced mainly through the diffusive shock acceleration mechanism operating at the strong shocks driven by supernova explosions. For the most part this will be the forward shock sweeping up the circumstellar material although there may be some particle acceleration at the reverse shock as it runs back into the ejecta (it should be noted that the total energy flux through the reverse shock is much less than that at the forward shock; in fact the time-integrated flux of energy through the forward shock can be several times the explosion energy as the PdV work done by the interior pressure is recycled through the shock during the remnant expansion).  The debate as to whether the GCR originate in superbubbles or SNRs seems to me to be a false dichotomy; most core-collapse SNe must occur spatially and temporally correlated with the birth of their progenitors in OB associations and thus in superbubbles, whereas the thermonuclear type 1a SNe are rather randomly distributed throughout the Galaxy, but surely both contribute to the GCR.

We would of course still like to have direct evidence, ideally an irrefutable "smoking gun" proof, that the GCR originate from SNRs \cite{1994A&A...287..959D, 2017MNRAS.471..201C}.  While we are not quite there yet substantial progress has been made in the last decade.  The current generation of imaging atmospheric Cherenkov telescopes  have conclusively shown that in at least some SNRs, charged particles are accelerated to $100\,\rm Tev$ and that the acceleration is linked to the outer shock, see e.g. the recent detailed study by the H.E.S.S. collaboration of the shell-type remnant RXJ1713.7-3946 \cite{2016arXiv160908671H} (see fig 2) which also shows some evidence for particles escaping from the shock into the surrounding medium \cite{2011MNRAS.415.1807D}. Unfortunately disentangling the relative contributions of electrons and nuclei has proven remarkably difficult, and the reality must be that they are both accelerated.  Indeed once they are relativistic the rest mass becomes unimportant and the only difference between protons and electrons as far as acceleration is concerned is the sign of the charge and the much stronger radiative losses for electrons.  The detection of what looks like the kinematic threshold for pion production by the Agile \cite{2011ApJ...742L..30G} and Fermi-Lat collaborations \cite{2013Sci...339..807A, 2016ApJ...816..100J} in the gamma-ray spectra of some SNRs is strong evidence for the acceleration of relativistic nuclei to at least GeV energies in these sources.  The non-thermal synchrotron X-ray emission observed in young remnants also supports the idea of magnetic field amplification, which in turn requires dynamically significant cosmic ray production (cosmic ray pressure comparable to the ram pressure of the shock) if this is by the Bell or similar mechanism \cite{2012A&ARv..20...49V}.

\begin{figure}
\begin{center}
\includegraphics[width=\textwidth]{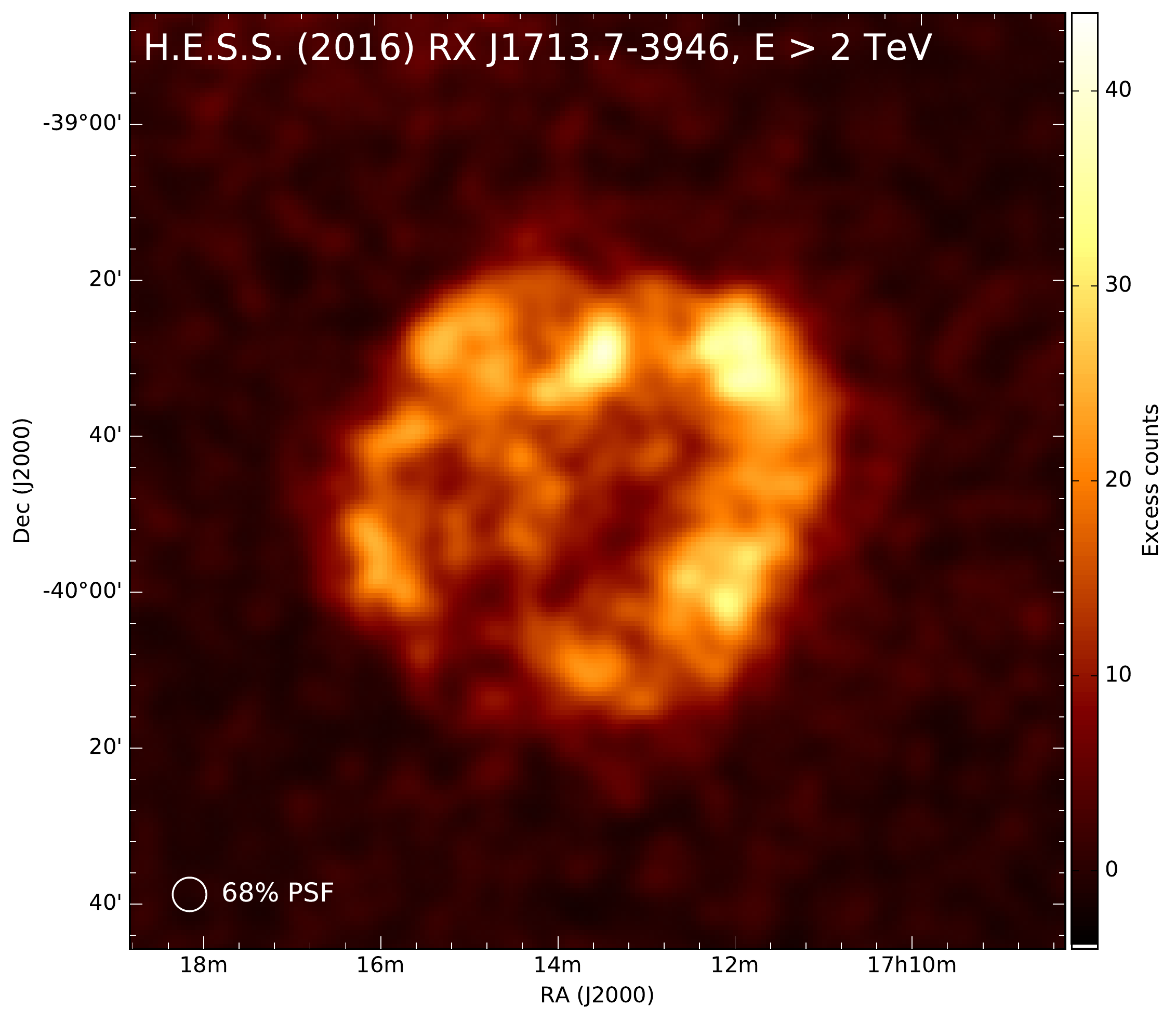}
\label{Fig 2}
\caption{The recent H.E.S.S. map of RXJ1713.7-3946 - a glowing shell of gamma-ray emission coincident with the outer shock of the SNR.}
\end{center}
\end{figure}

\section{Conclusions}

We live in exciting times.  The new precision data from the current generation of experiments is revealing fine detail in the GCR that clearly require a more quantitative and physical approach to interpretation and modelling.  We need to rethink the question of cosmic ray propagation in terms of a much more dynamic interstellar medium and Galaxy with nonlinear wave generation and cosmic ray pressure driven outflows (hopefully this will resolve the tension between current propagation models favouring soft production spectra and acceleration models favouring harder spectra).  We need to understand the formation of the Fermi bubbles and the possible role of Seyfert-like activity in the Galactic centre.  We need to improve our models of particle injection and chemical fractionation at shocks and see whether this can explain the compositional data and the discrepancy between the proton and helium spectra.  In direct detection we can already anticipate beautiful new data from experiments such as DAMPE and ISS-CREAM as well as continued and improved results from AMS02.  And in indirect methods the interpretation of air shower data is going to be much cleaner with the improvement in the hadronic models as a result of new LHC measurements.  We can anticipate roughly an order of magnitude improvement in the TeV gamma-ray observations from CTA, and continued progress in neutrino astronomy.  

The origin of cosmic rays remains, as foreseen by Ginzburg and Syrovatskii, a fundamental issue in the high-energy astrophysics of our Galaxy and continues to tantalise us, but we are making good progress.

\bibliographystyle{plain}
\bibliography{Luke}

\begin{thebibliography}{10}

\bibitem{2008PhRvL.101v1101A}
A.~A. {Abdo}, B.~{Allen}, T.~{Aune}, D.~{Berley}, E.~{Blaufuss}, S.~{Casanova},
  C.~{Chen}, B.~L. {Dingus}, R.~W. {Ellsworth}, L.~{Fleysher}, R.~{Fleysher},
  M.~M. {Gonzalez}, J.~A. {Goodman}, C.~M. {Hoffman}, P.~H. {H{\"u}ntemeyer},
  B.~E. {Kolterman}, C.~P. {Lansdell}, J.~T. {Linnemann}, J.~E. {McEnery},
  A.~I. {Mincer}, P.~{Nemethy}, D.~{Noyes}, J.~{Pretz}, J.~M. {Ryan}, P.~M.~S.
  {Parkinson}, A.~{Shoup}, G.~{Sinnis}, A.~J. {Smith}, G.~W. {Sullivan},
  V.~{Vasileiou}, G.~P. {Walker}, D.~A. {Williams}, and G.~B. {Yodh}.
\newblock {Discovery of Localized Regions of Excess 10-TeV Cosmic Rays}.
\newblock {\em Physical Review Letters}, 101(22):221101, November 2008.

\bibitem{2013Sci...339..807A}
M.~{Ackermann}, M.~{Ajello}, A.~{Allafort}, L.~{Baldini}, J.~{Ballet},
  G.~{Barbiellini}, M.~G. {Baring}, D.~{Bastieri}, K.~{Bechtol},
  R.~{Bellazzini}, R.~D. {Blandford}, E.~D. {Bloom}, E.~{Bonamente}, A.~W.
  {Borgland}, E.~{Bottacini}, T.~J. {Brandt}, J.~{Bregeon}, M.~{Brigida},
  P.~{Bruel}, R.~{Buehler}, G.~{Busetto}, S.~{Buson}, G.~A. {Caliandro}, R.~A.
  {Cameron}, P.~A. {Caraveo}, J.~M. {Casandjian}, C.~{Cecchi}, {\"O}.~{{\c
  C}elik}, E.~{Charles}, S.~{Chaty}, R.~C.~G. {Chaves}, A.~{Chekhtman}, C.~C.
  {Cheung}, J.~{Chiang}, G.~{Chiaro}, A.~N. {Cillis}, S.~{Ciprini}, R.~{Claus},
  J.~{Cohen-Tanugi}, L.~R. {Cominsky}, J.~{Conrad}, S.~{Corbel}, S.~{Cutini},
  F.~{D'Ammando}, A.~{de Angelis}, F.~{de Palma}, C.~D. {Dermer}, E.~{do Couto
  e Silva}, P.~S. {Drell}, A.~{Drlica-Wagner}, L.~{Falletti}, C.~{Favuzzi},
  E.~C. {Ferrara}, A.~{Franckowiak}, Y.~{Fukazawa}, S.~{Funk}, P.~{Fusco},
  F.~{Gargano}, S.~{Germani}, N.~{Giglietto}, P.~{Giommi}, F.~{Giordano},
  M.~{Giroletti}, T.~{Glanzman}, G.~{Godfrey}, I.~A. {Grenier}, M.-H.
  {Grondin}, J.~E. {Grove}, S.~{Guiriec}, D.~{Hadasch}, Y.~{Hanabata}, A.~K.
  {Harding}, M.~{Hayashida}, K.~{Hayashi}, E.~{Hays}, J.~W. {Hewitt}, A.~B.
  {Hill}, R.~E. {Hughes}, M.~S. {Jackson}, T.~{Jogler}, G.~{J{\'o}hannesson},
  A.~S. {Johnson}, T.~{Kamae}, J.~{Kataoka}, J.~{Katsuta}, J.~{Kn{\"o}dlseder},
  M.~{Kuss}, J.~{Lande}, S.~{Larsson}, L.~{Latronico}, M.~{Lemoine-Goumard},
  F.~{Longo}, F.~{Loparco}, M.~N. {Lovellette}, P.~{Lubrano}, G.~M. {Madejski},
  F.~{Massaro}, M.~{Mayer}, M.~N. {Mazziotta}, J.~E. {McEnery}, J.~{Mehault},
  P.~F. {Michelson}, R.~P. {Mignani}, W.~{Mitthumsiri}, T.~{Mizuno}, A.~A.
  {Moiseev}, M.~E. {Monzani}, A.~{Morselli}, I.~V. {Moskalenko}, S.~{Murgia},
  T.~{Nakamori}, R.~{Nemmen}, E.~{Nuss}, M.~{Ohno}, T.~{Ohsugi}, N.~{Omodei},
  M.~{Orienti}, E.~{Orlando}, J.~F. {Ormes}, D.~{Paneque}, J.~S. {Perkins},
  M.~{Pesce-Rollins}, F.~{Piron}, G.~{Pivato}, S.~{Rain{\`o}}, R.~{Rando},
  M.~{Razzano}, S.~{Razzaque}, A.~{Reimer}, O.~{Reimer}, S.~{Ritz},
  C.~{Romoli}, M.~{S{\'a}nchez-Conde}, A.~{Schulz}, C.~{Sgr{\`o}}, P.~E.
  {Simeon}, E.~J. {Siskind}, D.~A. {Smith}, G.~{Spandre}, P.~{Spinelli}, F.~W.
  {Stecker}, A.~W. {Strong}, D.~J. {Suson}, H.~{Tajima}, H.~{Takahashi},
  T.~{Takahashi}, T.~{Tanaka}, J.~G. {Thayer}, J.~B. {Thayer}, D.~J.
  {Thompson}, S.~E. {Thorsett}, L.~{Tibaldo}, O.~{Tibolla}, M.~{Tinivella},
  E.~{Troja}, Y.~{Uchiyama}, T.~L. {Usher}, J.~{Vandenbroucke}, V.~{Vasileiou},
  G.~{Vianello}, V.~{Vitale}, A.~P. {Waite}, M.~{Werner}, B.~L. {Winer}, K.~S.
  {Wood}, M.~{Wood}, R.~{Yamazaki}, Z.~{Yang}, and S.~{Zimmer}.
\newblock {Detection of the Characteristic Pion-Decay Signature in Supernova
  Remnants}.
\newblock {\em Science}, 339:807--811, February 2013.

\bibitem{2014ApJ...793...64A}
M.~{Ackermann}, A.~{Albert}, W.~B. {Atwood}, L.~{Baldini}, J.~{Ballet},
  G.~{Barbiellini}, D.~{Bastieri}, R.~{Bellazzini}, E.~{Bissaldi}, R.~D.
  {Blandford}, E.~D. {Bloom}, E.~{Bottacini}, T.~J. {Brandt}, J.~{Bregeon},
  P.~{Bruel}, R.~{Buehler}, S.~{Buson}, G.~A. {Caliandro}, R.~A. {Cameron},
  M.~{Caragiulo}, P.~A. {Caraveo}, E.~{Cavazzuti}, C.~{Cecchi}, E.~{Charles},
  A.~{Chekhtman}, J.~{Chiang}, G.~{Chiaro}, S.~{Ciprini}, R.~{Claus},
  J.~{Cohen-Tanugi}, J.~{Conrad}, S.~{Cutini}, F.~{D'Ammando}, A.~{de Angelis},
  F.~{de Palma}, C.~D. {Dermer}, S.~W. {Digel}, L.~{Di Venere}, E.~d.~C.~e.
  {Silva}, P.~S. {Drell}, C.~{Favuzzi}, E.~C. {Ferrara}, W.~B. {Focke},
  A.~{Franckowiak}, Y.~{Fukazawa}, S.~{Funk}, P.~{Fusco}, F.~{Gargano},
  D.~{Gasparrini}, S.~{Germani}, N.~{Giglietto}, F.~{Giordano}, M.~{Giroletti},
  G.~{Godfrey}, G.~A. {Gomez-Vargas}, I.~A. {Grenier}, S.~{Guiriec},
  D.~{Hadasch}, A.~K. {Harding}, E.~{Hays}, J.~W. {Hewitt}, X.~{Hou},
  T.~{Jogler}, G.~{J{\'o}hannesson}, A.~S. {Johnson}, W.~N. {Johnson},
  T.~{Kamae}, J.~{Kataoka}, J.~{Kn{\"o}dlseder}, D.~{Kocevski}, M.~{Kuss},
  S.~{Larsson}, L.~{Latronico}, F.~{Longo}, F.~{Loparco}, M.~N. {Lovellette},
  P.~{Lubrano}, D.~{Malyshev}, A.~{Manfreda}, F.~{Massaro}, M.~{Mayer}, M.~N.
  {Mazziotta}, J.~E. {McEnery}, P.~F. {Michelson}, W.~{Mitthumsiri},
  T.~{Mizuno}, M.~E. {Monzani}, A.~{Morselli}, I.~V. {Moskalenko}, S.~{Murgia},
  R.~{Nemmen}, E.~{Nuss}, T.~{Ohsugi}, N.~{Omodei}, M.~{Orienti}, E.~{Orlando},
  J.~F. {Ormes}, D.~{Paneque}, J.~H. {Panetta}, J.~S. {Perkins},
  M.~{Pesce-Rollins}, V.~{Petrosian}, F.~{Piron}, G.~{Pivato}, S.~{Rain{\`o}},
  R.~{Rando}, M.~{Razzano}, S.~{Razzaque}, A.~{Reimer}, O.~{Reimer},
  M.~{S{\'a}nchez-Conde}, M.~{Schaal}, A.~{Schulz}, C.~{Sgr{\`o}}, E.~J.
  {Siskind}, G.~{Spandre}, P.~{Spinelli}, {\L}.~{Stawarz}, A.~W. {Strong},
  D.~J. {Suson}, M.~{Tahara}, H.~{Takahashi}, J.~B. {Thayer}, L.~{Tibaldo},
  M.~{Tinivella}, D.~F. {Torres}, G.~{Tosti}, E.~{Troja}, Y.~{Uchiyama},
  G.~{Vianello}, M.~{Werner}, B.~L. {Winer}, K.~S. {Wood}, M.~{Wood}, and
  G.~{Zaharijas}.
\newblock {The Spectrum and Morphology of the Fermi Bubbles}.
\newblock {\em \apj}, 793:64, September 2014.

\bibitem{2011Sci...332...69A}
O.~{Adriani}, G.~C. {Barbarino}, G.~A. {Bazilevskaya}, R.~{Bellotti},
  M.~{Boezio}, E.~A. {Bogomolov}, L.~{Bonechi}, M.~{Bongi}, V.~{Bonvicini},
  S.~{Borisov}, S.~{Bottai}, A.~{Bruno}, F.~{Cafagna}, D.~{Campana},
  R.~{Carbone}, P.~{Carlson}, M.~{Casolino}, G.~{Castellini}, L.~{Consiglio},
  M.~P. {De Pascale}, C.~{De Santis}, N.~{De Simone}, V.~{Di Felice}, A.~M.
  {Galper}, W.~{Gillard}, L.~{Grishantseva}, G.~{Jerse}, A.~V. {Karelin}, S.~V.
  {Koldashov}, S.~Y. {Krutkov}, A.~N. {Kvashnin}, A.~{Leonov}, V.~{Malakhov},
  V.~{Malvezzi}, L.~{Marcelli}, A.~G. {Mayorov}, W.~{Menn}, V.~V. {Mikhailov},
  E.~{Mocchiutti}, A.~{Monaco}, N.~{Mori}, N.~{Nikonov}, G.~{Osteria},
  F.~{Palma}, P.~{Papini}, M.~{Pearce}, P.~{Picozza}, C.~{Pizzolotto},
  M.~{Ricci}, S.~B. {Ricciarini}, L.~{Rossetto}, R.~{Sarkar}, M.~{Simon},
  R.~{Sparvoli}, P.~{Spillantini}, Y.~I. {Stozhkov}, A.~{Vacchi},
  E.~{Vannuccini}, G.~{Vasilyev}, S.~A. {Voronov}, Y.~T. {Yurkin}, J.~{Wu},
  G.~{Zampa}, N.~{Zampa}, and V.~G. {Zverev}.
\newblock {PAMELA Measurements of Cosmic-Ray Proton and Helium Spectra}.
\newblock {\em Science}, 332:69, April 2011.

\bibitem{2015PhRvL.115u1101A}
M.~{Aguilar}, D.~{Aisa}, B.~{Alpat}, A.~{Alvino}, G.~{Ambrosi}, K.~{Andeen},
  L.~{Arruda}, N.~{Attig}, P.~{Azzarello}, A.~{Bachlechner}, and et~al.
\newblock {Precision Measurement of the Helium Flux in Primary Cosmic Rays of
  Rigidities 1.9 GV to 3 TV with the Alpha Magnetic Spectrometer on the
  International Space Station}.
\newblock {\em Physical Review Letters}, 115(21):211101, November 2015.

\bibitem{2015PhRvL.114q1103A}
M.~{Aguilar}, D.~{Aisa}, B.~{Alpat}, A.~{Alvino}, G.~{Ambrosi}, K.~{Andeen},
  L.~{Arruda}, N.~{Attig}, P.~{Azzarello}, A.~{Bachlechner}, and et~al.
\newblock {Precision Measurement of the Proton Flux in Primary Cosmic Rays from
  Rigidity 1 GV to 1.8 TV with the Alpha Magnetic Spectrometer on the
  International Space Station}.
\newblock {\em Physical Review Letters}, 114(17):171103, May 2015.

\bibitem{2016PhRvL.117o1103A}
M.~{Ahlers}.
\newblock {Deciphering the Dipole Anisotropy of Galactic Cosmic Rays}.
\newblock {\em Physical Review Letters}, 117(15):151103, October 2016.

\bibitem{2017PrPNP..94..184A}
M.~{Ahlers} and P.~{Mertsch}.
\newblock {Origin of small-scale anisotropies in Galactic cosmic rays}.
\newblock {\em Progress in Particle and Nuclear Physics}, 94:184--216, May
  2017.

\bibitem{2010ApJ...714L..89A}
H.~S. {Ahn}, P.~{Allison}, M.~G. {Bagliesi}, J.~J. {Beatty}, G.~{Bigongiari},
  J.~T. {Childers}, N.~B. {Conklin}, S.~{Coutu}, M.~A. {DuVernois}, O.~{Ganel},
  J.~H. {Han}, J.~A. {Jeon}, K.~C. {Kim}, M.~H. {Lee}, L.~{Lutz}, P.~{Maestro},
  A.~{Malinin}, P.~S. {Marrocchesi}, S.~{Minnick}, S.~I. {Mognet}, J.~{Nam},
  S.~{Nam}, S.~L. {Nutter}, I.~H. {Park}, N.~H. {Park}, E.~S. {Seo}, R.~{Sina},
  J.~{Wu}, J.~{Yang}, Y.~S. {Yoon}, R.~{Zei}, and S.~Y. {Zinn}.
\newblock {Discrepant Hardening Observed in Cosmic-ray Elemental Spectra}.
\newblock {\em \apjl}, 714:L89--L93, May 2010.

\bibitem{2006AdSpR..37.1950A}
H.~S. {Ahn}, E.~S. {Seo}, J.~H. {Adams}, G.~{Bashindzhagyan}, K.~E. {Batkov},
  J.~{Chang}, M.~{Christl}, A.~R. {Fazely}, O.~{Ganel}, R.~M. {Gunasingha},
  T.~G. {Guzik}, J.~{Isbert}, K.~C. {Kim}, E.~{Kouznetsov}, M.~{Panasyuk},
  A.~{Panov}, W.~K.~H. {Schmidt}, R.~{Sina}, N.~V. {Sokolskaya}, J.~Z. {Wang},
  J.~P. {Wefel}, J.~{Wu}, and V.~{Zatsepin}.
\newblock {The energy spectra of protons and helium measured with the ATIC
  experiment}.
\newblock {\em Advances in Space Research}, 37:1950--1954, 2006.

\bibitem{1950PhRv...77..375A}
H.~{Alfv{\'e}n}.
\newblock {On the Solar Origin of Cosmic Radiation. II}.
\newblock {\em Physical Review}, 77:375--379, February 1950.

\bibitem{1949PhRv...75..892A}
H.~{Alfven}, R.~D. {Richtmyer}, and E.~{Teller}.
\newblock {On the Origin of Cosmic Rays}.
\newblock {\em Physical Review}, 75:892--893, March 1949.

\bibitem{2015A&A...583A..95A}
R.~{Aloisio}, P.~{Blasi}, and P.~D. {Serpico}.
\newblock {Nonlinear cosmic ray Galactic transport in the light of AMS-02 and
  Voyager data}.
\newblock {\em \aap}, 583:A95, November 2015.

\bibitem{1977ICRC...11..132A}
W.~I. {Axford}, E.~{Leer}, and G.~{Skadron}.
\newblock {The acceleration of cosmic rays by shock waves}.
\newblock {\em International Cosmic Ray Conference}, 11:132--137, 1977.

\bibitem{1934PhRv...46...76B}
W.~{Baade} and F.~{Zwicky}.
\newblock {Remarks on Super-Novae and Cosmic Rays}.
\newblock {\em Physical Review}, 46:76--77, July 1934.

\bibitem{1978MNRAS.182..147B}
A.~R. {Bell}.
\newblock {The acceleration of cosmic rays in shock fronts. I}.
\newblock {\em \mnras}, 182:147--156, January 1978.

\bibitem{2004MNRAS.353..550B}
A.~R. {Bell}.
\newblock {Turbulent amplification of magnetic field and diffusive shock
  acceleration of cosmic rays}.
\newblock {\em \mnras}, 353:550--558, September 2004.

\bibitem{2017APS..APR.M4009B}
W.~R. {Binns}, M.~H. {Israel}, E.~R. {Christian}, A.~C. {Cummings}, G.~A. {de
  Nolfo}, K.~A. {Lave}, R.~A. {Leske}, R.~A. {Mewaldt}, E.~C. {Stone}, T.~T.
  {von Rosenvinge}, and M.~E. {Wiedenbeck}.
\newblock {Observation of the $^{60}$Fe Nucleosynthesis-Clock Isotope in
  Galactic Cosmic Rays}.
\newblock In {\em APS April Meeting Abstracts}, January 2017.

\bibitem{1978ApJ...221L..29B}
R.~D. {Blandford} and J.~P. {Ostriker}.
\newblock {Particle acceleration by astrophysical shocks}.
\newblock {\em \apjl}, 221:L29--L32, April 1978.

\bibitem{1971CoASP...3..140B}
G.~{Burbidge} and K.~{Brecher}.
\newblock {Extragalactic Cosmic Rays-A Reappraisal}.
\newblock {\em Comments on Astrophysics and Space Physics}, 3:140, September
  1971.

\bibitem{1964PPS....84..141B}
G.~R. {Burbidge} and F.~{Hoyle}.
\newblock {On cosmic rays as an extragalactic phenomenon}.
\newblock {\em Proceedings of the Physical Society}, 84:141--150, July 1964.

\bibitem{2014ApJ...783...91C}
D.~{Caprioli} and A.~{Spitkovsky}.
\newblock {Simulations of Ion Acceleration at Non-relativistic Shocks. I.
  Acceleration Efficiency}.
\newblock {\em \apj}, 783:91, March 2014.

\bibitem{2014ApJ...794...46C}
D.~{Caprioli} and A.~{Spitkovsky}.
\newblock {Simulations of Ion Acceleration at Non-relativistic Shocks. II.
  Magnetic Field Amplification}.
\newblock {\em \apj}, 794:46, October 2014.

\bibitem{2014ApJ...794...47C}
D.~{Caprioli} and A.~{Spitkovsky}.
\newblock {Simulations of Ion Acceleration at Non-relativistic Shocks. III.
  Particle Diffusion}.
\newblock {\em \apj}, 794:47, October 2014.

\bibitem{2017arXiv170408252C}
D.~{Caprioli}, D.~T. {Yi}, and A.~{Spitkovsky}.
\newblock {Chemical Enhancements in Shock-accelerated Particles: Ab-initio
  Simulations}.
\newblock {\em ArXiv e-prints}, April 2017.

\bibitem{1982ApJ...258..860C}
M.~{Casse} and J.~A. {Paul}.
\newblock {On the stellar origin of the Ne-22 excess in cosmic rays}.
\newblock {\em \apj}, 258:860--863, July 1982.

\bibitem{1981IAUS...94..361C}
C.~J. {Cesarsky} and J.-P. {Bibring}.
\newblock {Cosmic-ray injection into shock-waves}.
\newblock In G.~{Setti}, G.~{Spada}, and A.~W. {Wolfendale}, editors, {\em
  Origin of Cosmic Rays}, volume~94 of {\em IAU Symposium}, page 361, 1981.

\bibitem{1981ICRC....2..269C}
C.~J. {Cesarsky}, R.~{Rothenflug}, and M.~{Casse}.
\newblock {Mass per charge ratio in hot plasmas and cosmic ray source
  composition}.
\newblock {\em International Cosmic Ray Conference}, 2:269--272, 1981.

\bibitem{2017MNRAS.471..201C}
P.~{Cristofari}, S.~{Gabici}, T.~B. {Humensky}, M.~{Santander}, R.~{Terrier},
  E.~{Parizot}, and S.~{Casanova}.
\newblock {Supernova remnants in the very-high-energy gamma-ray domain: the
  role of the Cherenkov telescope array}.
\newblock {\em \mnras}, 471:201--209, October 2017.

\bibitem{2015ApJ...808..107C}
R.~M. {Crocker}, G.~V. {Bicknell}, A.~M. {Taylor}, and E.~{Carretti}.
\newblock {A Unified Model of the Fermi Bubbles, Microwave Haze, and Polarized
  Radio Lobes: Reverse Shocks in the Galactic Center's Giant Outflows}.
\newblock {\em \apj}, 808:107, August 2015.

\bibitem{2014arXiv1407.2144D}
G.~{Di Sciascio} and R.~{Iuppa}.
\newblock {On the Observation of the Cosmic Ray Anisotropy below 10$^{15}$ eV}.
\newblock {\em ArXiv e-prints}, July 2014.

\bibitem{2012ApJ...747...40D}
J.~{Donnelly}, A.~{Thompson}, D.~{O'Sullivan}, J.~{Daly}, L.~{Drury},
  V.~{Domingo}, and K.-P. {Wenzel}.
\newblock {Actinide and Ultra-Heavy Abundances in the Local Galactic Cosmic
  Rays: An Analysis of the Results from the LDEF Ultra-Heavy Cosmic-Ray
  Experiment}.
\newblock {\em \apj}, 747:40, March 2012.

\bibitem{2014MNRAS.444..365D}
T.~P. {Downes} and L.~O. {Drury}.
\newblock {Cosmic ray pressure driven magnetic field amplification:
  dimensional, radiative and field orientation effects}.
\newblock {\em \mnras}, 444:365--375, October 2014.

\bibitem{1983RPPh...46..973D}
L.~O. {Drury}.
\newblock {An introduction to the theory of diffusive shock acceleration of
  energetic particles in tenuous plasmas}.
\newblock {\em Reports on Progress in Physics}, 46:973--1027, August 1983.

\bibitem{2011MNRAS.415.1807D}
L.~O. {Drury}.
\newblock {Escaping the accelerator: how, when and in what numbers do cosmic
  rays get out of supernova remnants?}
\newblock {\em \mnras}, 415:1807--1814, August 2011.

\bibitem{2017A&A...597A.117D}
L.~O.~'. {Drury} and A.~W. {Strong}.
\newblock {Power requirements for cosmic ray propagation models involving
  diffusive reacceleration; estimates and implications for the damping of
  interstellar turbulence}.
\newblock {\em \aap}, 597:A117, January 2017.

\bibitem{1994A&A...287..959D}
L.~O. {Drury}, F.~A. {Aharonian}, and H.~J. {Voelk}.
\newblock {The gamma-ray visibility of supernova remnants. A test of cosmic ray
  origin}.
\newblock {\em \aap}, 287:959--971, July 1994.

\bibitem{1989A&A...225..179D}
L.~O. {Drury}, W.~J. {Markiewicz}, and H.~J. {Voelk}.
\newblock {Simplified models for the evolution of supernova remnants including
  particle acceleration}.
\newblock {\em \aap}, 225:179--191, November 1989.

\bibitem{1997ApJ...487..197E}
D.~C. {Ellison}, L.~O. {Drury}, and J.-P. {Meyer}.
\newblock {Galactic Cosmic Rays from Supernova Remnants. II. Shock Acceleration
  of Gas and Dust}.
\newblock {\em \apj}, 487:197--217, September 1997.

\bibitem{1980MNRAS.193..723E}
R.~I. {Epstein}.
\newblock {The acceleration of interstellar grains and the composition of the
  cosmic rays}.
\newblock {\em \mnras}, 193:723--729, December 1980.

\bibitem{2017PhRvL.119c1101G}
D.~{Gaggero}, D.~{Grasso}, A.~{Marinelli}, M.~{Taoso}, and A.~{Urbano}.
\newblock {Diffuse Cosmic Rays Shining in the Galactic Center: A Novel
  Interpretation of H.E.S.S. and Fermi-LAT {$\gamma$} -Ray Data}.
\newblock {\em Physical Review Letters}, 119(3):031101, July 2017.

\bibitem{2017arXiv170609812G}
Y.~{Genolini}, P.~D. {Serpico}, M.~{Boudaud}, S.~{Caroff}, V.~{Poulin},
  L.~{Derome}, J.~{Lavalle}, D.~{Maurin}, V.~{Poireau}, S.~{Rosier},
  P.~{Salati}, and M.~{Vecchi}.
\newblock {Indications for a high-rigidity break in the cosmic-ray diffusion
  coefficient}.
\newblock {\em ArXiv e-prints}, June 2017.

\bibitem{2017arXiv170201001G}
G.~{Giacinti} and J.~G. {Kirk}.
\newblock {Cosmic-Ray Anisotropy and the Local Interstellar Turbulence}.
\newblock {\em ArXiv e-prints}, February 2017.

\bibitem{2017ApJ...835..258G}
G.~{Giacinti} and J.~G. {Kirk}.
\newblock {Large-scale Cosmic-Ray Anisotropy as a Probe of Interstellar
  Turbulence}.
\newblock {\em \apj}, 835:258, February 2017.

\bibitem{2012PhRvL.109g1101G}
G.~{Giacinti} and G.~{Sigl}.
\newblock {Local Magnetic Turbulence and TeV-PeV Cosmic Ray Anisotropies}.
\newblock {\em Physical Review Letters}, 109(7):071101, August 2012.

\bibitem{2009ApJ...692.1075G}
S.~{Gillessen}, F.~{Eisenhauer}, S.~{Trippe}, T.~{Alexander}, R.~{Genzel},
  F.~{Martins}, and T.~{Ott}.
\newblock {Monitoring Stellar Orbits Around the Massive Black Hole in the
  Galactic Center}.
\newblock {\em \apj}, 692:1075--1109, February 2009.

\bibitem{1972NPhS..239....8G}
V.~L. {Ginzburg}.
\newblock {Gamma Radiation of Magellanic Clouds and Metagalactic Origin of
  Cosmic Rays}.
\newblock {\em Nature Physical Science}, 239:8--9, September 1972.

\bibitem{1964ocr..book.....G}
V.~L. {Ginzburg} and S.~I. {Syrovatskii}.
\newblock {\em {The Origin of Cosmic Rays}}.
\newblock Pergamon Press, Oxford, authorised english translation edition, 1964.

\bibitem{2011ApJ...742L..30G}
A.~{Giuliani}, M.~{Cardillo}, M.~{Tavani}, Y.~{Fukui}, S.~{Yoshiike},
  K.~{Torii}, G.~{Dubner}, G.~{Castelletti}, G.~{Barbiellini}, A.~{Bulgarelli},
  P.~{Caraveo}, E.~{Costa}, P.~W. {Cattaneo}, A.~{Chen}, T.~{Contessi}, E.~{Del
  Monte}, I.~{Donnarumma}, Y.~{Evangelista}, M.~{Feroci}, F.~{Gianotti},
  F.~{Lazzarotto}, F.~{Lucarelli}, F.~{Longo}, M.~{Marisaldi}, S.~{Mereghetti},
  L.~{Pacciani}, A.~{Pellizzoni}, G.~{Piano}, P.~{Picozza}, C.~{Pittori},
  G.~{Pucella}, M.~{Rapisarda}, A.~{Rappoldi}, S.~{Sabatini}, P.~{Soffitta},
  E.~{Striani}, M.~{Trifoglio}, A.~{Trois}, S.~{Vercellone}, F.~{Verrecchia},
  V.~{Vittorini}, S.~{Colafrancesco}, P.~{Giommi}, and G.~{Bignami}.
\newblock {Neutral Pion Emission from Accelerated Protons in the Supernova
  Remnant W44}.
\newblock {\em \apjl}, 742:L30, December 2011.

\bibitem{2015ARA&A..53..199G}
I.~A. {Grenier}, J.~H. {Black}, and A.~W. {Strong}.
\newblock {The Nine Lives of Cosmic Rays in Galaxies}.
\newblock {\em \araa}, 53:199--246, August 2015.

\bibitem{1972spre.conf.1617G}
N.~L. {Grigorov}, V.~E. {Nesterov}, I.~D. {Rapoport}, and I.~A. {Savenko}.
\newblock {Study of energy spectra of primary cosmic rays at very high energies
  on the Proton series of satellites.}
\newblock In S.~A. {Bowhill}, L.~D. {Jaffe}, and M.~J. {Rycroft}, editors, {\em
  Space Research Conference}, volume~2 of {\em Space Research Conference},
  pages 1617--1622, 1972.

\bibitem{2016arXiv160908671H}
{H.~E.~S.~S.~Collaboration}, {:}, H.~{Abdalla}, H.~{Abdalla}, A.~{Abramowski},
  F.~{Aharonian}, F.~{Ait Benkhali}, A.~G. {Akhperjanian}, T.~{Andersson},
  E.~O. {Ang{\"u}ner}, and et~al.
\newblock {H.E.S.S. observations of RX J1713.7-3946 with improved angular and
  spectral resolution; evidence for gamma-ray emission extending beyond the
  X-ray emitting shell}.
\newblock {\em ArXiv e-prints}, September 2016.

\bibitem{2017arXiv170702744H}
A.~{Hanusch}, T.~{Liseykina}, and M.~{Malkov}.
\newblock {Anomalies in Cosmic Ray Composition: Explanation Based on Mass to
  Charge Ratio}.
\newblock {\em ArXiv e-prints}, July 2017.

\bibitem{2016Natur.531..476H}
{HESS Collaboration}, A.~{Abramowski}, F.~{Aharonian}, F.~A. {Benkhali}, A.~G.
  {Akhperjanian}, E.~O. {Ang{\"u}ner}, M.~{Backes}, A.~{Balzer},
  Y.~{Becherini}, J.~B. {Tjus}, and et~al.
\newblock {Acceleration of petaelectronvolt protons in the Galactic Centre}.
\newblock {\em \nat}, 531:476--479, March 2016.

\bibitem{2017NCimC..39..313I}
M.~{Incagli} and {AMS02 Collaboration}.
\newblock {AMS02 results after 4 years of data taking on the International
  Space Station}.
\newblock {\em Nuovo Cimento C Geophysics Space Physics C}, 39:313, July 2017.

\bibitem{2005NuPhA.758..201I}
M.~H. {Israel}, W.~R. {Binns}, A.~C. {Cummings}, R.~A. {Leske}, R.~A.
  {Mewaldt}, E.~C. {Stone}, T.~T. {von Rosenvinge}, and M.~E. {Wiedenbeck}.
\newblock {Isotopic Composition of Cosmic Rays: Results from the Cosmic Ray
  Isotope Spectrometer on the ACE Spacecraft}.
\newblock {\em Nuclear Physics A}, 758:201--208, July 2005.

\bibitem{2016ApJ...816..100J}
T.~{Jogler} and S.~{Funk}.
\newblock {Revealing W51C as a Cosmic Ray Source Using Fermi-LAT Data}.
\newblock {\em \apj}, 816:100, January 2016.

\bibitem{2016ApJ...824...16J}
G.~{J{\'o}hannesson}, R.~{Ruiz de Austri}, A.~C. {Vincent}, I.~V. {Moskalenko},
  E.~{Orlando}, T.~A. {Porter}, A.~W. {Strong}, R.~{Trotta}, F.~{Feroz},
  P.~{Graff}, and M.~P. {Hobson}.
\newblock {Bayesian Analysis of Cosmic Ray Propagation: Evidence against
  Homogeneous Diffusion}.
\newblock {\em \apj}, 824:16, June 2016.

\bibitem{1977DoSSR.234.1306K}
G.~F. {Krymskii}.
\newblock {A regular mechanism for the acceleration of charged particles on the
  front of a shock wave}.
\newblock {\em Akademiia Nauk SSSR Doklady}, 234:1306--1308, June 1977.

\bibitem{1983A&A...125..249L}
P.~O. {Lagage} and C.~J. {Cesarsky}.
\newblock {The maximum energy of cosmic rays accelerated by supernova shocks}.
\newblock {\em \aap}, 125:249--257, September 1983.

\bibitem{2017PhRvD..95f3009L}
P.~{Lipari}.
\newblock {Interpretation of the cosmic ray positron and antiproton fluxes}.
\newblock {\em \prd}, 95(6):063009, March 2017.

\bibitem{1981heaa.book.....L}
M.~S. {Longair}.
\newblock {\em {High energy astrophysics}}.
\newblock Canbridge University Press, 1st edition, 1981.

\bibitem{2000MNRAS.314...65L}
S.~G. {Lucek} and A.~R. {Bell}.
\newblock {Non-linear amplification of a magnetic field driven by cosmic ray
  streaming}.
\newblock {\em \mnras}, 314:65--74, May 2000.

\bibitem{2014NIMPA.742..241M}
S.~M. {Mari} and P.~{Montini}.
\newblock {Energy spectrum of cosmic ray protons and helium nuclei measured by
  the ARGO-YBJ experiment}.
\newblock {\em Nuclear Instruments and Methods in Physics Research A},
  742:241--244, April 2014.

\bibitem{1997ApJ...487..182M}
J.-P. {Meyer}, L.~O. {Drury}, and D.~C. {Ellison}.
\newblock {Galactic Cosmic Rays from Supernova Remnants. I. A Cosmic-Ray
  Composition Controlled by Volatility and Mass-to-Charge Ratio}.
\newblock {\em \apj}, 487:182--196, September 1997.

\bibitem{2016ApJ...831..148M}
R.~P. {Murphy}, M.~{Sasaki}, W.~R. {Binns}, T.~J. {Brandt}, T.~{Hams}, M.~H.
  {Israel}, A.~W. {Labrador}, J.~T. {Link}, R.~A. {Mewaldt}, J.~W. {Mitchell},
  B.~F. {Rauch}, K.~{Sakai}, E.~C. {Stone}, C.~J. {Waddington}, N.~E. {Walsh},
  J.~E. {Ward}, and M.~E. {Wiedenbeck}.
\newblock {Galactic Cosmic Ray Origins and OB Associations: Evidence from
  SuperTIGER Observations of Elements $_{26}$Fe through $_{40}$Zr}.
\newblock {\em \apj}, 831:148, November 2016.

\bibitem{2016A&A...588A..86N}
A.~{Neronov} and G.~{Meynet}.
\newblock {Abundances of $^{59}$Co and $^{59}$Ni in the cosmic ray flux}.
\newblock {\em \aap}, 588:A86, April 2016.

\bibitem{2013arXiv1305.6752O}
L.~{O'C.~Drury}.
\newblock {The problem of small angular scale structure in the cosmic ray
  anisotropy data}.
\newblock {\em ArXiv e-prints}, May 2013.

\bibitem{2006astro.ph.12377P}
A.~D. {Panov}, J.~H. {Adams}, H.~S. {Ahn}, G.~L. {Bashindzhagyan}, K.~E.
  {Batkov}, J.~{Chang}, M.~{Christl}, A.~R. {Fazely}, O.~{Ganel}, R.~M.
  {Gunashingha}, T.~G. {Guzik}, J.~{Isbert}, K.~C. {Kim}, E.~N. {Kouznetsov},
  M.~I. {Panasyuk}, W.~K.~H. {Schmidt}, E.~S. {Seo}, N.~V. {Sokolskaya}, J.~W.
  {Watts}, J.~P. {Wefel}, J.~{Wu}, and V.~I. {Zatsepin}.
\newblock {The results of ATIC-2 experiment for elemental spectra of cosmic
  rays}.
\newblock {\em ArXiv Astrophysics e-prints}, December 2006.

\bibitem{2012A&A...538A..80P}
N.~{Prantzos}.
\newblock {On the origin and composition of Galactic cosmic rays}.
\newblock {\em \aap}, 538:A80, February 2012.

\bibitem{2009ApJ...697.2083R}
B.~F. {Rauch}, J.~T. {Link}, K.~{Lodders}, M.~H. {Israel}, L.~M. {Barbier},
  W.~R. {Binns}, E.~R. {Christian}, J.~R. {Cummings}, G.~A. {de Nolfo},
  S.~{Geier}, R.~A. {Mewaldt}, J.~W. {Mitchell}, S.~M. {Schindler}, L.~M.
  {Scott}, E.~C. {Stone}, R.~E. {Streitmatter}, C.~J. {Waddington}, and M.~E.
  {Wiedenbeck}.
\newblock {Cosmic Ray origin in OB Associations and Preferential Acceleration
  of Refractory Elements: Evidence from Abundances of Elements $_{26}$Fe
  through $_{34}$Se}.
\newblock {\em \apj}, 697:2083--2088, June 2009.

\bibitem{2017MNRAS.470..865R}
S.~{Recchia}, P.~{Blasi}, and G.~{Morlino}.
\newblock {Cosmic ray-driven winds in the Galactic environment and the cosmic
  ray spectrum}.
\newblock {\em \mnras}, 470:865--881, September 2017.

\bibitem{2013Sci...341..150S}
E.~C. {Stone}, A.~C. {Cummings}, F.~B. {McDonald}, B.~C. {Heikkila}, N.~{Lal},
  and W.~R. {Webber}.
\newblock {Voyager 1 Observes Low-Energy Galactic Cosmic Rays in a Region
  Depleted of Heliospheric Ions}.
\newblock {\em Science}, 341:150--153, July 2013.

\bibitem{2017arXiv170803005T}
{The HAWC Collaboration} and {The IceCube Collaboration}.
\newblock {Combined Analysis of Cosmic-Ray Anisotropy with IceCube and HAWC}.
\newblock {\em ArXiv e-prints}, August 2017.

\bibitem{2012A&ARv..20...49V}
J.~{Vink}.
\newblock {Supernova remnants: the X-ray perspective}.
\newblock {\em \aapr}, 20:49, December 2012.

\bibitem{2015ApJ...815..119V}
E.~E. {Vos} and M.~S. {Potgieter}.
\newblock {New Modeling of Galactic Proton Modulation during the Minimum of
  Solar Cycle 23/24}.
\newblock {\em \apj}, 815:119, December 2015.

\bibitem{2016MNRAS.463.2893W}
M.~{Walls}, M.~{Chernyakova}, R.~{Terrier}, and A.~{Goldwurm}.
\newblock {Examining molecular clouds in the Galactic Centre region using X-ray
  reflection spectra simulations}.
\newblock {\em \mnras}, 463:2893--2903, December 2016.

\bibitem{1998NIMPB.145..409W}
B.~A. {Weaver}, A.~J. {Westphal}, P.~B. {Price}, V.~G. {Afanasyev}, and V.~V.
  {Akimov}.
\newblock {Performance of the Ultraheavy Collector of the Trek experiment}.
\newblock {\em Nuclear Instruments and Methods in Physics Research B},
  145:409--428, November 1998.

\bibitem{1999ApJ...523L..61W}
M.~E. {Wiedenbeck}, W.~R. {Binns}, E.~R. {Christian}, A.~C. {Cummings}, B.~L.
  {Dougherty}, P.~L. {Hink}, J.~{Klarmann}, R.~A. {Leske}, M.~{Lijowski}, R.~A.
  {Mewaldt}, E.~C. {Stone}, M.~R. {Thayer}, T.~T. {von Rosenvinge}, and N.~E.
  {Yanasak}.
\newblock {Constraints on the Time Delay between Nucleosynthesis and Cosmic-Ray
  Acceleration from Observations of $^{59}$Ni and $^{59}$Co}.
\newblock {\em \apjl}, 523:L61--L64, September 1999.

\bibitem{2011ApJ...728..122Y}
Y.~S. {Yoon}, H.~S. {Ahn}, P.~S. {Allison}, M.~G. {Bagliesi}, J.~J. {Beatty},
  G.~{Bigongiari}, P.~J. {Boyle}, J.~T. {Childers}, N.~B. {Conklin},
  S.~{Coutu}, M.~A. {DuVernois}, O.~{Ganel}, J.~H. {Han}, J.~A. {Jeon}, K.~C.
  {Kim}, M.~H. {Lee}, L.~{Lutz}, P.~{Maestro}, A.~{Malinine}, P.~S.
  {Marrocchesi}, S.~A. {Minnick}, S.~I. {Mognet}, S.~{Nam}, S.~{Nutter}, I.~H.
  {Park}, N.~H. {Park}, E.~S. {Seo}, R.~{Sina}, S.~{Swordy}, S.~P. {Wakely},
  J.~{Wu}, J.~{Yang}, R.~{Zei}, and S.~Y. {Zinn}.
\newblock {Cosmic-ray Proton and Helium Spectra from the First CREAM Flight}.
\newblock {\em \apj}, 728:122, February 2011.

\bibitem{2016JPhCS.767a2027Z}
M.~{Zhang} and N.~{Pogorelov}.
\newblock {The Heliosphere as Seen in TeV Cosmic Rays}.
\newblock In {\em Journal of Physics Conference Series}, volume 767 of {\em
  Journal of Physics Conference Series}, page 012027, November 2016.

\bibitem{1996A&A...311..113Z}
V.~N. {Zirakashvili}, D.~{Breitschwerdt}, V.~S. {Ptuskin}, and H.~J. {Voelk}.
\newblock {Magnetohydrodynamic wind driven by cosmic rays in a rotating
  galaxy.}
\newblock {\em \aap}, 311:113--126, July 1996.

\end{thebibliography}

\end{document}